\documentclass[%
unsortedaddress,
nofootinbib,
 amsmath,amssymb,
 aps,
twocolumn,
]{revtex4}
\usepackage[utf8]{inputenc}
\usepackage[english]{babel}
\usepackage{titlesec}
\usepackage{etoolbox}

\usepackage{amsmath}
\usepackage{amssymb} 

\usepackage{graphicx} 
\usepackage{xcolor}
\usepackage{natbib}
\usepackage{braket}

\newcommand{\dd}{\mathrm{d}}

\newcommand{\inv}{^{-1}}

\usepackage[normalem]{ulem}
\newcommand{\stkout}[1]{\ifmmode\text{\sout{\ensuremath{#1}}}\else\sout{#1}\fi}

\usepackage[all]{nowidow}

\begin{document}

\preprint{APS/123-QED}

\title{Conditioning Normalizing Flows for Rare Event Sampling}
\date{\today}

\author{Sebastian Falkner 
}
\affiliation{University of Vienna, Faculty of Physics \& Vienna Doctoral School in Physics, 1090 Vienna, Austria}

\author{Alessandro Coretti 
}
\affiliation{University of Vienna, Faculty of Physics, 1090 Vienna, Austria.}

\author{Salvatore Romano
}
\affiliation{University of Vienna, Faculty of Physics \& Vienna Doctoral School in Physics, 1090 Vienna, Austria}

\author{Phillip Geissler 
}
\thanks{Phillip L. Geissler passed away on July 17, 2022, while this paper was being completed.}
\affiliation{Department of Chemistry, University of California, Berkeley, California 94720, USA}

\author{Christoph Dellago 
}
\email{christoph.dellago@univie.ac.at}
\affiliation{University of Vienna, Faculty of Physics, 1090 Vienna, Austria.}

\begin{abstract}

Understanding the dynamics of complex molecular processes is often linked to the study of infrequent transitions between long-lived stable states. The standard approach to the sampling of such rare events is to generate an ensemble of transition paths using a random walk in trajectory space. This, however, comes with the drawback of strong correlations between subsequently sampled paths and with an intrinsic difficulty in parallelizing the sampling process. We propose a transition path sampling scheme based on neural-network generated configurations. These are obtained employing normalizing flows, a neural network class able to generate statistically independent samples from a given distribution. With this approach, not only are correlations between visited paths removed, but the sampling process becomes easily parallelizable. Moreover, by conditioning the normalizing flow, the sampling of configurations can be steered towards regions of interest. We show that this approach enables the resolution of both the thermodynamics and kinetics of the transition region.

\end{abstract}

\maketitle

\section{Introduction}

The exponential increase in computational power experienced by computers since the advent of molecular simulations has radically changed basically all aspects of the study of statistical mechanics via numerical experiments. Simulations of rare events have also benefited from such improvements, but progress in this area has relied even more on methodological developments rather than the exploitation of raw computing power. This is due to the intrinsic nature of rare events, which are phenomena that occur infrequently, but happen quickly if they occur. The resulting time scale disparity is often so large that such processes cannot be simulated even on the fastest computers with straightforward methods. Examples are omnipresent in physics, chemistry and biology and include nucleation processes~\cite{Menzl2016a,Arjun2019}, protein folding~\cite{Juraszek2006,Okazaki2019}, dynamics of ions in solution~\cite{Geissler1999,Ballard2012,Schwierz2020,Falkner2021} and chemical reactions~\cite{Geissler2001,Leitold2020}. All of these processes exhibit transitions between stable states separated by high energetic and/or entropic barriers. Resolving the thermodynamics and kinetics at the barrier top is the key challenge for understanding the rare event.

Over the years, many enhanced sampling methods were developed to focus the computational effort on regions of interest in phase space. For instance, when one aims to resolve the thermodynamic properties of a system, umbrella sampling~\cite{TorrieG.MValleau1977} received widespread recognition. In this approach, a harmonic bias is added to the potential energy function, efficiently restricting the sampling to certain regions of configuration space. In contrast, when investigating the kinetics of a rare event, a properly weighted set of unbiased reactive trajectories is desired. Transition path sampling (TPS)~\cite{Bolhuis2002} is an efficient strategy to achieve this goal by performing a Markov chain Monte Carlo simulation in trajectory space. A basic scheme for generating a new path based on a previous one is the shooting move~\cite{Csajka1998}, where a point on the previous trajectory is randomly selected, possibly perturbed, and then integrated forward and backward in time until a stable state is reached. If the newly generated trajectory connects the stable states, it is accepted and used for the generation of the next path. 

Despite improvements introduced by different shooting schemes~\cite{Peters2006,Brotzakis2016,Jung2017}, the foundation of these sampling approaches is the generation of a new path from the previous one. Therefore, these algorithms are inherently sequential and correlations between subsequently visited paths are inevitable. Even though a high acceptance rate may be achieved, a strong similarity between subsequent paths degrades the efficiency of sampling. 

With recent developments in the field of generative neural networks~\cite{NIPS2014_5ca3e9b1,Kingma2014,Papamakarios2019}, the sampling of independent equilibrium configurations from the Boltzmann distribution came into reach. In particular, normalizing flows~\cite{Papamakarios2019} have already been applied to free energy calculations~\cite{Wirnsberger2020}, exploration of configuration space~\cite{Noe2019,Wirnsberger2022}, finding minimum energy paths~\cite{liu2022pathflow} and force field parametrization~\cite{Koehler2022}. In this work, we propose a parallel sampling scheme to explore the reactive path space based on normalizing flows in the form of Boltzmann generators~\cite{Noe2019} for the generation of shooting points. The flow model is conditioned to steer the generation to regions of interest, which at the same time allows for the accurate reconstruction of free energy profiles.

\begin{figure*}
    \centering
    \includegraphics[width=17.4cm]{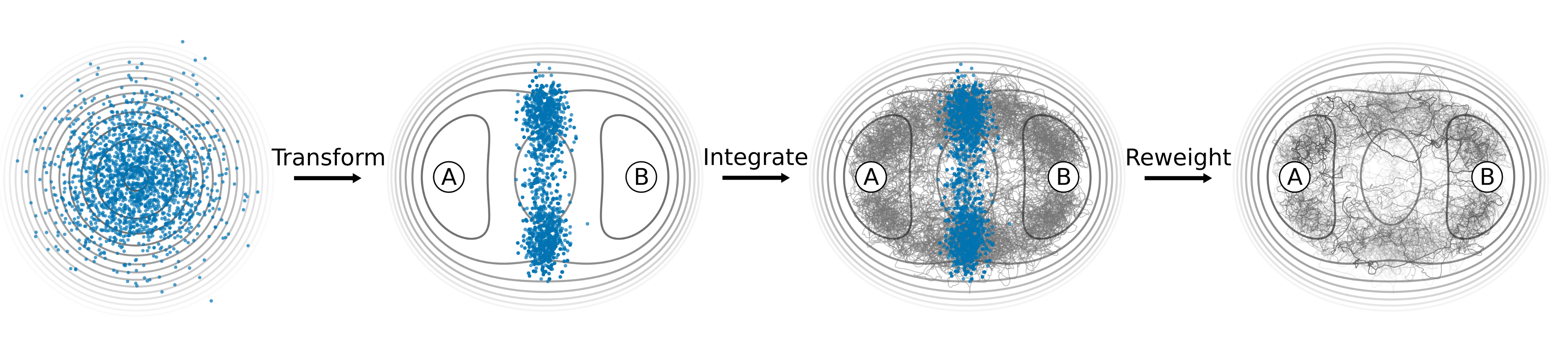}
    \caption{\textbf{Schematic overview of the parallel path sampling algorithm starting from neural network-generated shooting points.} From left to right: (1) Sampling from the Gaussian latent space, (2) transformation into the shooting point distribution via a neural network, (3) integration of the equations of motion forward and backward in time, (4) reweighting of transition paths to obtain an unbiased ensemble.}
    \label{fig:parallel_sampling_scheme}
\end{figure*}

The proposed path sampling scheme starts by sampling points from a multivariate Gaussian distribution. These points are then transformed into shooting points using a conditioned Boltzmann generator. From these, trajectories are obtained by integrating forward and backward in time until a stable state is reached. The resulting paths are reweighted to obtain a properly weighted transition path ensemble. In the following, we will describe this algorithm in detail and demonstrate it using some illustrative models.

\subsection{Flexible Length Transition Path Sampling}
The path ensemble targeted by our sampling procedure includes all reactive trajectories connecting stable states. Each trajectory is defined as a sequence of configurations $X(\tau) = \left\{x_{0}, x_{\Delta t}, x_{2\Delta t}, ..., x_\tau \right\}$, where $\tau$ is the length of the path and is a multiple of the timestep $\Delta t$. Transition paths connect two given stable states, A and B, and they are required to have exactly one point in each of these states. Consequently, transition pathways have varying lengths $\tau$.

Transition paths are sampled proportional to their statistical weight $P_{\text{AB}}\left[ X(\tau) \right]$. Here we consider the probabilities within a small region $\text{d}X^\tau$ in path space. Accordingly, the probability of a reactive path $X(\tau)$ can be expressed as:
\begin{align}
    \label{reactive_path_probability}
    P_{\text{AB}} \left[ X(\tau) \right] \text{d}X^\tau =
    &\frac{1}{Z_{\text{AB}}}\,
    H_{\text{AB}}(x_0, x_\tau)  \notag \\
    &\times \prod_{i=1}^{\tau/\Delta t-1} \widetilde{h}(x_{i \Delta t})\ 
    P\left[ X(\tau) \right] \text{d}X^\tau ,
\end{align}
where $H_{\text{AB}}(x_0, x_\tau)$ is unity if the trajectory connects states A and B in any order and is zero otherwise. More explicitly, this function is defined as
\begin{equation}
 H_{\text{AB}}(x_0, x_\tau) = 
\begin{cases}
1 & \text{if} \ h_{\text{A}}(x_0) h_{\text{B}}(x_\tau) = 1\ \\ &\text{or} \  h_{\text{A}}(x_\tau)h_{\text{B}}(x_0)  = 1 \vspace{0.15cm} ;\\
0 & \text{otherwise} .
\end{cases}
\end{equation}
Here, $h_{\text{A}}$ and $h_{\text{B}}$ are population functions that return one if a point lies in state A and B, respectively, and vanish otherwise. The function $\widetilde{h}(x)$ is defined as:
\begin{equation}
\widetilde{h}(x) = 
\begin{cases}
0 & \text{if} \ h_{\text{A}}(x) = 1\ \text{or} \ h_{\text{B}}(x) = 1 ;\\
1 & \text{otherwise} ,
\end{cases}
\end{equation}
and acts as a constraint to focus attention only on values of $\tau$ that are comparable to a natural transition time. The normalizing factor $Z_{\text{AB}}$ has the form of a partition function:
\begin{align}
    \label{reactive_path_partition_function}
    Z_{\text{AB}} = \sum_{\tau} \int \text{d}X^\tau\,H_{\text{AB}}(x_0, x_\tau) \prod_{i=1}^{\tau/\Delta t-1} \widetilde{h}(x_{i \Delta t})\, P\left[ X(\tau) \right] .
\end{align}
Assuming Markovian dynamics, the dynamical path probability $P\left[ X(\tau) \right] \text{d}X^\tau$ is defined based on the equilibrium probability of the starting point $p_{\text{eq}}(x_0)$ and the short-time transition probabilities $p(x_{i\Delta t} \to x_{(i+1)\Delta t})$:
\begin{align}
    \label{dynamical_path_probability}
    P\left[ X(\tau) \right]\, \text{d}X^\tau = p_{\text{eq}}(x_0) \prod_{i=0}^{\tau/\Delta t-1} p(x_{i\Delta t} \to x_{(i+1)\Delta t}) \, \text{d}X^\tau .
\end{align}

\subsection{Parallel Path Sampling}
Shooting moves are an integral part of most path sampling schemes. Their efficiency relies on the fact that shooting points in a region of high $p\left(\text{TP} | x \right)$, which is the probability of generating a transition path given a certain configuration $x$, lead to an efficient exploration of the path ensemble. However, the scalability of TPS with shooting moves is limited by the inherently sequential nature of the sampling. The previous trajectory is indispensable for the generation of the new trajectory. To combine the efficiency of shooting moves with the possibility of parallel sampling, we propose an alternative algorithm to sample the transition path ensemble.

The basis of the scheme is a set of shooting points generated before the actual path sampling starts. These configurations can be sampled from an arbitrary distribution denoted as $p_{\text{SP}}(x)$. From these shooting points, trajectories are obtained by integration forward and backward in time until a stable state is reached. As a result, the generation of paths becomes embarrassingly parallel because the trajectories are generated independently from each other. Fleeting trajectories initiated using configurations from $p_{\text{SP}}(x)$, however, do not correspond to a properly weighted transition path ensemble. One difference are the missing population functions to distinguish paths that connect stable states from ones that end in the same state in both directions. Even more critically, paths that dwell a long time in high probability regions of $p_{\text{SP}}(x)$ are sampled preferentially. Accordingly, a reweighting factor $w\left[X(\tau) \right]$ has to be included when calculating the expectation values of path observables:
\begin{align}
\label{reweighted_observable}
\langle A\left[X(\tau)\right] \rangle \approx
    \frac{\sum_{i=1}^{N} w\left[ X(\tau)_i \right] A\left[X(\tau)_i \right]}{\sum_{i=1}^{N} w\left[ X(\tau)_i \right] } .
\end{align}
A similar path reweighting has already been successfully applied in other studies, e.g. in works by Daru et al.~\cite{Daru2014} and Menzl et al.~\cite{Menzl2016b} for the calculation of rate constants.

To derive an expression for $w\left[ X(\tau)_i \right]$ in the context of our parallel sampling scheme, we first consider the generation probability of a trajectory obtained from an \textit{a priori} sampled shooting point. By means of a shooting move, a given trajectory can be generated from any of its points. Therefore, the total generation probability is the sum of the independent generation probabilities from each point on the trajectory:

\begin{align}
\label{generation_probability_from_psp}
P_{\text{gen}}\left[ X(\tau) \right]\,& \text{d}X^{\tau} = \text{d}X^{\tau}\,
    \sum_{k=0}^{\tau/\Delta t} \Bigg[ p_{\text{SP}}(x_{k\Delta t})
    \frac{p_{\text{eq}}(x_{0})}{p_{\text{eq}}(x_{k\Delta t})}\notag \\
    &\times \bigl[h_{\text{A}}(x_0) + h_{\text{B}}(x_0)\bigr] \bigl[h_{\text{A}}(x_\tau) + h_{\text{B}}(x_\tau)\bigr]\notag\\
    &\times \prod_{j=1}^{\tau/\Delta t-1} \widetilde{h}(x_{j \Delta t})
    \prod_{i=0}^{\tau/\Delta t-1} p(x_{i\Delta t} \to x_{(i+1)\Delta t})  \Bigg] .
\end{align} 

The weight of reactive paths in Eq.~\ref{generation_probability_from_psp} differs from the corresponding weight in the transition path ensemble (Eq.~\ref{reactive_path_probability}) by the factor:
\begin{align}
\label{reweighting_factor_path}
w \left[ X(\tau) \right] = 
    \frac{1}
    {Z_{\text{AB}}} 
    \left[\sum_{k=0}^{\tau/\Delta t} \frac{p_{\text{SP}}(x_{k\Delta t}) }{p_{\text{eq}}(x_{k\Delta t})} \right]^{-1} .
\end{align} 
A full derivation of the generation probability and of the reweighting factor is given in the SI.
Since we are solely interested in properly weighting a path relative to all others, the partition function $Z_{\text{AB}}$ can be omitted. This leads to a tractable relative reweighting factor to recover a properly weighted transition path ensemble given a collection of trajectories generated from a distribution of shooting points.

In the simplest case, one can choose the equilibrium distribution $p_{\text{eq}}(x)$ as the shooting point distribution so that $p_{\text{SP}}(x) \equiv p_{\text{eq}}(x)$ with the caveat that points already lying in a stable state must be sorted out. The reweighting factor then reduces to $(\tau/\Delta t + 1)^{-1} / Z_{\text{AB}}$. In this case, reactive paths are weighted by their inverse number of points. When all shooting points lie on a dividing surface and have weights according to the equilibrium distribution, the above reweighting factor reduces to the inverse number of crossings of the path with the surface, which agrees with the findings of Best and Hummer~\cite{Hummer2004, Best2005}. Given an infinite number of samples from $p_{\text{SP}}(x)$ as a set of shooting points, ergodicity in transition path space is guaranteed if every configuration with a non-zero probability in $p_{\text{eq}}(x)$ also has a non-zero probability in $p_{\text{SP}}(x)$.

\subsection{Targeted Sampling using Boltzmann Generators}
Enhanced sampling revolves around the efficient exploration of low probability regions in configuration space. In standard equilibrium simulations, these regions are often not visited frequently enough to make accurate predictions about the thermodynamics of the system. Therefore, one common approach in enhanced sampling methods is to restrict the sampling to the region of interest and thereby focus computational resources. Often this is achieved by applying a bias towards the region of interest. For example, in umbrella sampling this bias is introduced in the form of a harmonic bias potential, which is added to the potential energy function of the system. This bias is defined using a collective variable, here denoted as $r(x)$, a bias center $\bar{r}$ and a force constant $k$:
\begin{align}
U^{\text{bias}}(x,\bar{r}) = U(x) + \frac{k}{2} \left[ r(x) - \bar{r} \right]^2 .
\end{align}

By the addition of the bias potential, configurations with a collective variable value close to $\bar{r}$ will be sampled more often. The resulting configuration ensemble is referred to as an umbrella window. In the canonical ensemble, the probability of observing a configuration $x$ given an applied bias potential centered at $\bar{r}$ can be expressed as:
\begin{align}
\label{biased_sample_probabilty}
p_{\text{biased}}(x | \bar{r}) &= 
    \frac{ 1 }{ Z_\text{x} } \exp \bigl\{ -\beta [ U(x) + \frac{k}{2} (r(x) - \bar{r})^2] \bigr\},\\
Z_\text{x} &= \int\dd x \ \exp \bigl\{-\beta [ U(x) + \frac{k}{2} (r(x) - \bar{r})^2]\bigr\} .
\end{align}

Boltzmann generators, as proposed by No\'e and coworkers~\cite{Noe2019}, provide a way to obtain uncorrelated samples from such a distribution. They belong to the class of flow-based generative models and they allow to obtain unbiased samples from a given target distribution. This unique feature makes them well-suited for our parallel path sampling scheme. 

In flow-based models, a neural network learns an invertible coordinate transformation between an easy to sample latent distribution $p_\text{z}(z)$ and a complex data distribution $p_\text{x}(x)$:
\begin{align}
	x & = f(z; \theta) \label{eq:f_dir} ,\\
	z & = f\inv(x; \theta) \label{eq:f_inv} ,
\end{align}
where $x$ and $z$ represent samples from the data space (denoted as a whole by $\text{x}$) and from the latent space (denoted as a whole by $\text{z}$), respectively, while $\theta$ represents the set of trainable network parameters that parameterize the transformation $f$.

\begin{figure}
    \centering
    \includegraphics[width=0.9\columnwidth]{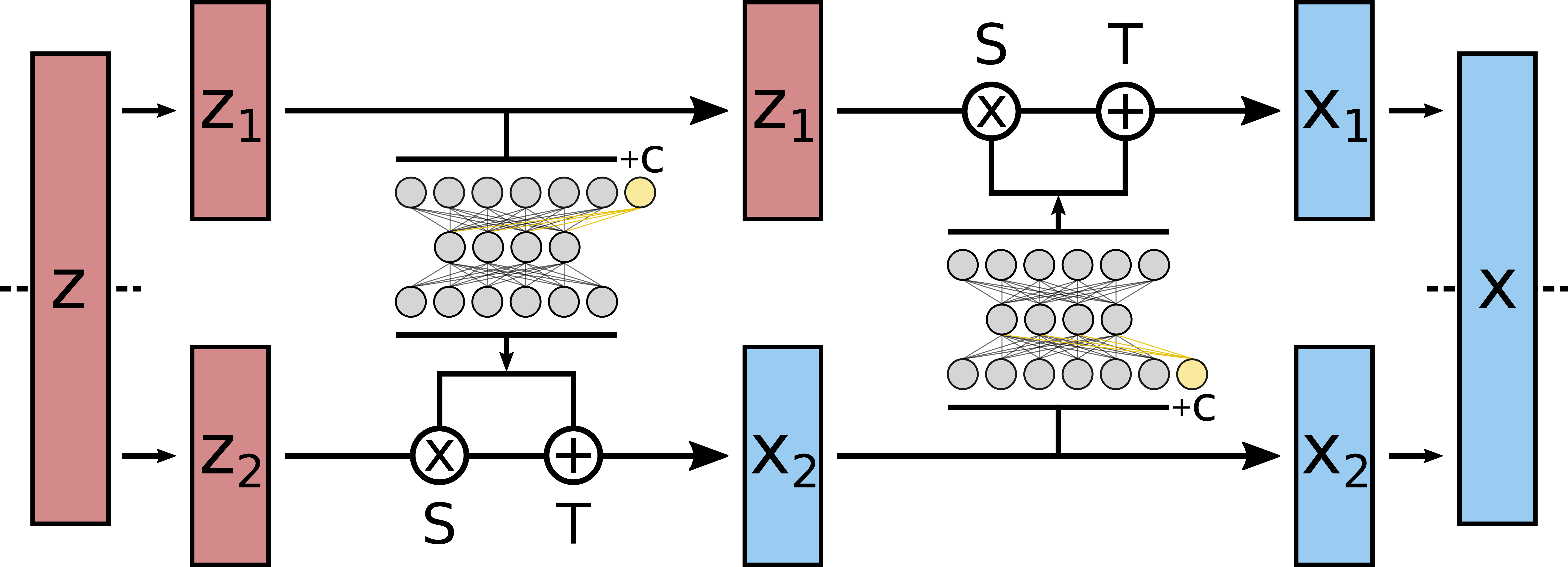}
    \caption{\textbf{Schematic overview of the split coupling flow architecture used in this work.} The input $z$ (red) is split in two parts $z_{1}$ and $z_{2}$. An identity transformation is applied to $z_1$. Using $z_1$ as an input to a feed-forward neural network (gray), scaling and shifting parameters $S$ and $T$ for the transformation of $z_2$ to $x_2$ (blue) are obtained. Subsequently the process is repeated in the other direction to obtain the fully transformed output $x$. If a conditioned transformation is desired, the condition $c$ is appended to the input of the feed-forward layer (yellow).  }
    \label{fig:network_architecture}
\end{figure}

The architecture of Boltzmann generators is based on a split-coupling flow using RealNVP blocks as proposed by Dinh et al.~\cite{Dinh2017}. Split-coupling flows allow to compute the determinant of the transformation's Jacobian efficiently~\cite{Dinh2017}. Therefore, the distribution of neural-network generated samples can be expressed via the change of variable theorem:
\begin{align}
    \label{change_of_variable}
	q_\text{x}(x) &= p_\text{z}\left[ f\inv(x; \theta) \right]\ \lvert \det J_{f\inv}(x; \theta) \rvert , \\
	q_\text{z}(z) &= p_\text{x}\left[ f(z; \theta) \right]\ \lvert \det J_{f}(z; \theta) \rvert ,
\end{align}
where $q$ represents the distributions generated by the network in the corresponding spaces, which will be, in general, different from $p$. The determinant of the Jacobian is tractable thanks to the particular construction of the network, as shown schematically in Fig.~\ref{fig:network_architecture}. The input is split into two parts, $x_1$ and $x_2$. While an identity transformation is applied to one part, the other one is scaled and translated with parameters that are a function of the first. The generated distribution $q_\text{x}(x)$ is, in general, only an approximation to the Boltzmann distribution. However, since the probability of a generated sample can be obtained using Eq.~\ref{change_of_variable}, a statistical weight can be assigned to each generated configuration in order to correct for the bias and to recover the exact distribution. A simple choice for this reweighting factor~\cite{Noe2019} is represented by the ratio between the reference and generated probability.

The generation of samples with probabilities according to Eq.~\ref{biased_sample_probabilty} for a single bias center is possible in the framework of Boltzmann generators. 
However, it is rarely the case that a single window provides sufficient information on the rare event of interest. For this reason, the generation of samples should be possible at arbitrary bias centers.

Conditioning the transformation applied by the normalizing flow enables sampling at different bias centers using a single neural network. A simple scheme to condition a split-coupling flow architecture was proposed by Ardizzone and coworkers~\cite{Ardizzone2019} in relation to image generation. Here, the transformation is conditioned by concatenating the condition data $c$ to the coupling layer network input, as indicated in Fig.~\ref{fig:network_architecture}. For the purpose of generating configurations at different bias centers with weights given by Eq.~\ref{biased_sample_probabilty}, the condition vector $c$ corresponds to the bias center $\bar{r}$. This approach leaves the latent space distribution unconditioned and it imposes the condition directly on the transformation which is then reflected on the generated distribution. The change of variable theorem then takes the form
\begin{align}
	q_\text{x}(x | \bar{r}) &= p_\text{z}\left[ f\inv(x | \bar{r}; \theta) \right]\ \lvert \det J\inv(x | \bar{r}; \theta) \rvert , \\
	q_\text{z}(z | \bar{r}) &= p_\text{x}\left[ f(z | \bar{r}; \theta) | \bar{r} \right]\ \lvert \det J(z | \bar{r}; \theta) \rvert .
\end{align}

In Boltzmann generators, the goal is to learn a transformation between samples obtained from a Gaussian and from the Boltzmann distribution. Thanks to the invertibility of the transformation, training of the generator can be performed in both directions, i.e. from Gaussian to Boltzmann and vice versa. The training loss function is formulated based on the Kullback–Leibler (KL) divergence between the generated and reference distributions. Conditioning of the transformation can then be incorporated in the definition of the loss functions for the training. In \emph{training by example}, samples from different umbrella windows are transformed into Gaussian-distributed samples. Here, the training loss $L_\text{fwd}$ is given by the conditional KL-divergence between the reference and generated data distribution $\text{KL}\left[p_\text{x}(x | r ) || q_\text{x}(x | r; \theta)\right]$ (full derivation in SI):

\begin{align}
\label{conditional_example_loss}
L_\text{fwd} = \mathbf{E}_{\bar{r} \sim p(r)} \biggl\{ \mathbf{E}_{x \sim  p_\text{x}(x | \bar{r})} &\biggl[\frac{1}{\sigma^2} || f\inv(x | \bar{r}; \theta) || ^2 \notag \\
&- \lvert \det J\inv(x | \bar{r}; \theta) \rvert \biggr] \biggr\} .
\end{align}
Here, $p(r)$ describes the arbitrary distribution of the condition variable. For conditioning on bias centers, a uniform distribution where inaccessible regions are masked is well-suited. In practice, configurations at discrete bias positions on the collective variable are sampled. These discrete positions should cover the regions of interest in $p(r)$. The samples are transformed and the parameters of the network are optimized with respect to Eq.~\ref{conditional_example_loss}.

\begin{figure}
    \centering
    \includegraphics[width=0.8\columnwidth]{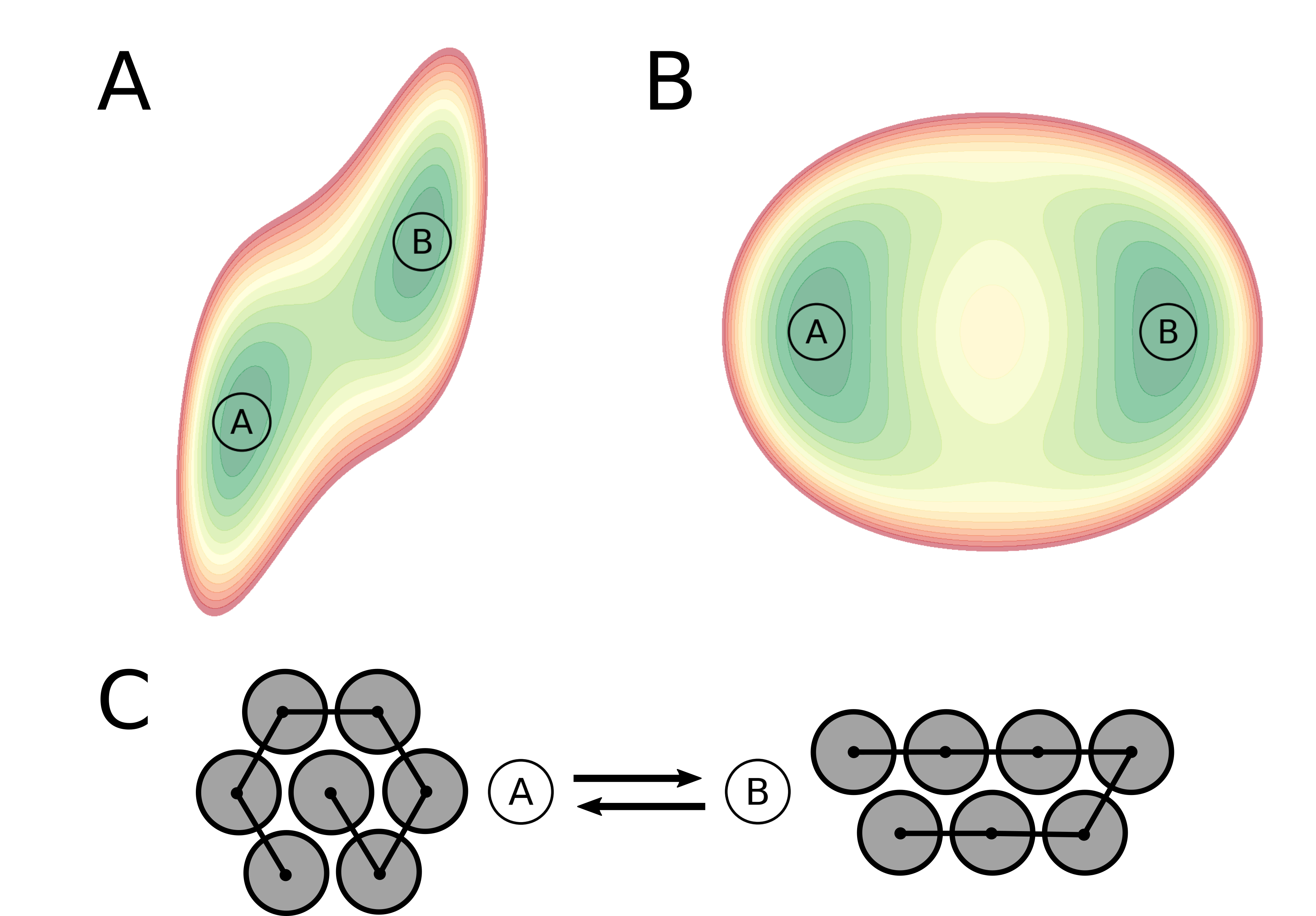}
    \caption{\textbf{Overview of the model systems including the state definitions.} The potential energy surface of the two-dimensional double well~\cite{Jung2017} (A) and the bistable double well model (B). For the polymer model (C), only the stable states are depicted.}
    \label{fig:model_systems}
\end{figure}

Training in the other direction, the \emph{training by energy}, works by sampling from the latent Gaussian distribution and transforming to the desired umbrella windows. The conditional KL-divergence between the reference and generated latent distribution $\text{KL}\left[p_\text{z}(z | r ) || q_\text{z}(z | r; \theta)\right] $ is then minimized leading to the loss function $L_\text{rev}$ (full derivation in SI):
\begin{align}
\label{conditional_energy_loss}
L_\text{rev} = &\mathbf{E}_{\bar{r} \sim p(r)}  \biggl\{ \mathbf{E}_{z \sim  p_\text{z}(z | \bar{r})} \biggl[\beta U(f(z | \bar{r}; \theta)) \notag\\
&+ \beta \frac{k}{2} [r(f(z | \bar{r}; \theta)) - \bar{r}]^2 - \lvert \det J(z | \bar{r}; \theta) \rvert \biggr] \biggr\} .
\end{align}
Consequently, training by energy is initiated by sampling from the distribution of bias centers and from the latent distribution. Latent points and corresponding bias centers are then transformed and parameters of the network are optimized with respect to Eq.~\ref{conditional_energy_loss}. During training by energy, the network can be trained at different temperatures by adjusting the variance of the Gaussian latent space distribution~\cite{Noe2019}. 
The final loss function for the training can be computed as 
\begin{equation}
    L = \lambda_{\text{fwd}}L_{\text{fwd}} + \lambda_{\text{rev}}L_{\text{rev}} ,
\end{equation}
where $\lambda_{\text{fwd}}$ and $\lambda_{\text{rev}}$ are weights used to tune the focus of the training.

\section{Results}

\subsection{Resolving the Barrier Region}

\begin{figure}
    \centering
    \includegraphics[width=0.9\columnwidth]{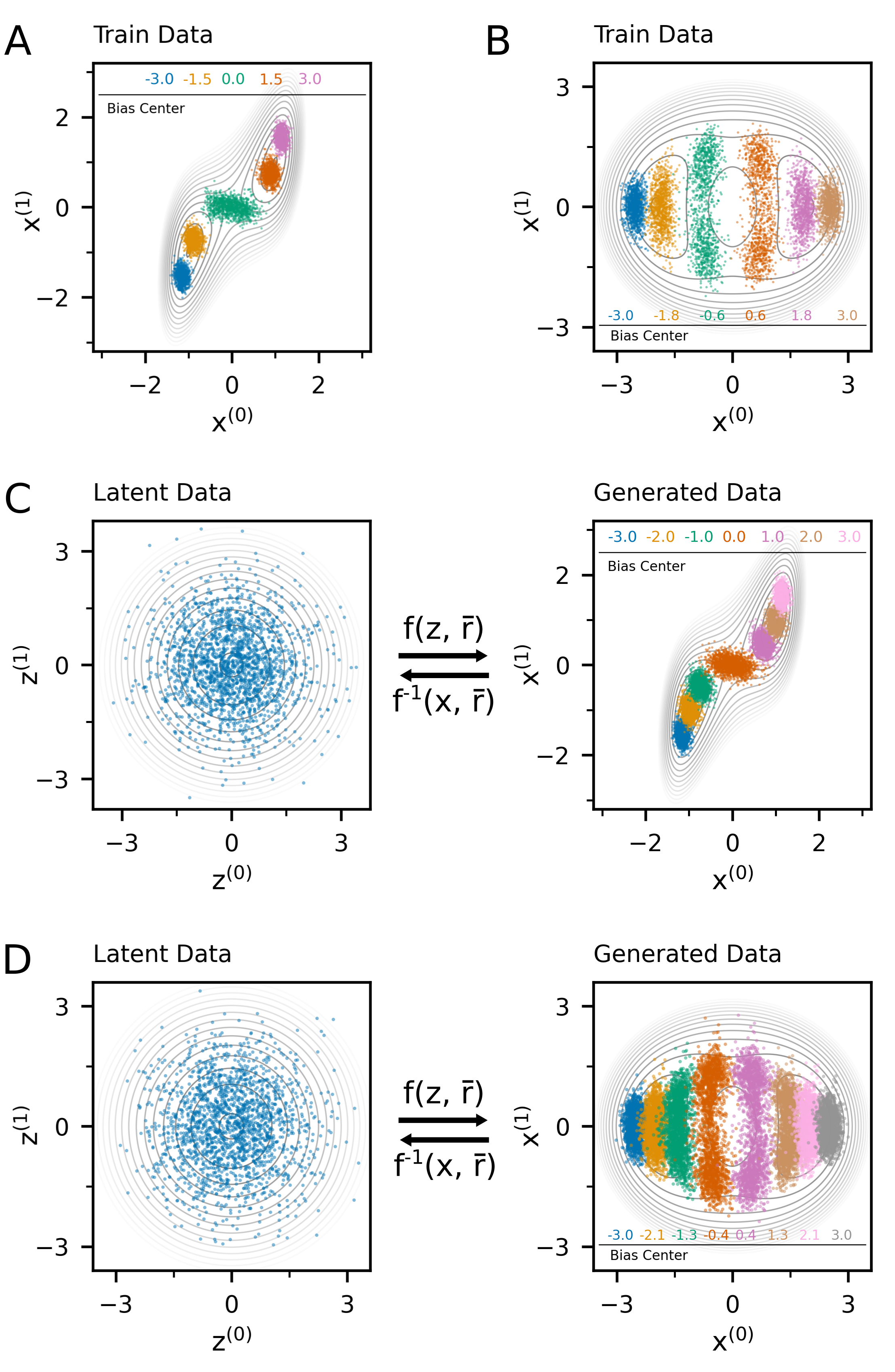}
    \caption{\textbf{Training configurations and network-generated configurations for the two-dimensional model systems.} (A, B) The training data consists of samples at different bias centers along the reaction coordinate. (C, D) Samples from a Gaussian (Latent Space, left) are transformed using the conditioned Boltzmann generator to obtain configurations at varying bias centers (Generated Data, right). }
    \label{fig:DW_data}
\end{figure}

We first test the conditioned Boltzmann generators on a simple two-dimensional model~\cite{Jung2017} (Fig.~\ref{fig:model_systems}A, system parameters in SI). Here, the conditioning of the Boltzmann generator greatly improves the resolution of low probability regions in configuration space, as shown in Fig.~\ref{fig:DW_data}. Analogous to umbrella sampling, a bias potential is applied to force the system to regions that are rarely seen in equilibrium at a given temperature. In the case of the original Boltzmann generator, low probability states can be included in the generated distribution by the introduction of a reaction coordinate loss~\cite{Noe2019}. Here, the entropy of samples projected on a reaction coordinate was maximized during training. While this is sufficient to encourage a broad sampling of the target distribution and to prevent a mode collapse, it does not allow a targeted sampling of low probability regions. For an accurate free energy estimate and especially for the generation of transition states, the sampling of specific low probability regions in configuration space must be enhanced. Due to the conditioning of the transformation, the generator can be steered to focus on certain regions in configurations space, see Fig.~\ref{fig:DW_data}. 

\begin{figure}
    \centering
    \includegraphics[width=0.7\columnwidth]{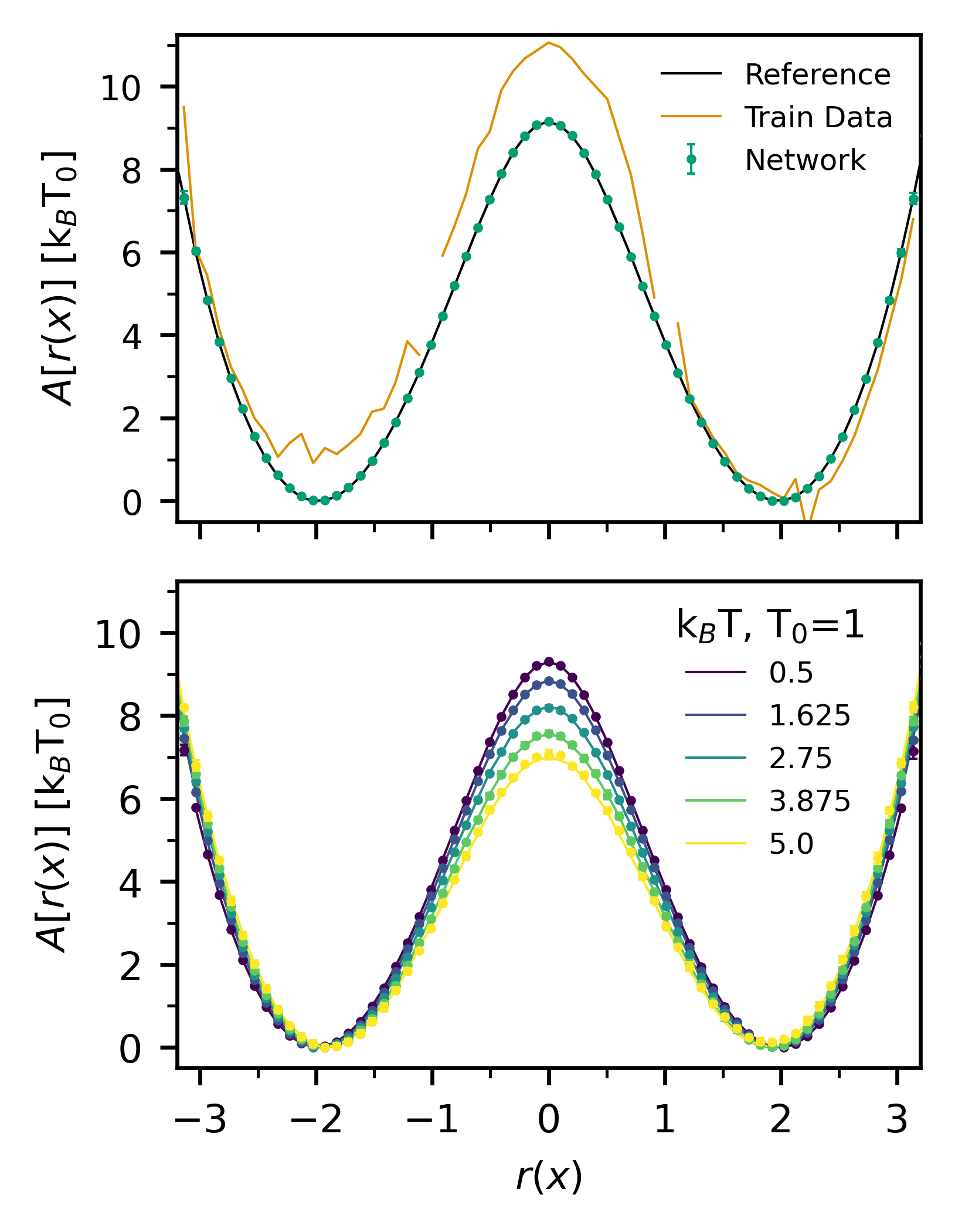}
    \caption{\textbf{Free energy reconstruction from network-generated configurations for the double well model.} The upper panel shows the free energy profile at constant temperature estimated using reference data, training data (see figure \ref{fig:DW_data}A) and network-generated configurations. In the lower panel, the same network was used to estimate free energy profiles at different temperatures. A reference profile is shown for each temperature as a solid line, the profiles obtained from network-generated configurations are shown as points. 
    }
    \label{fig:DW_free_energy}
\end{figure}

From network-generated configurations at different bias centers, accurate free energy profiles can be reconstructed. Using the weighted histogram analysis method~\cite{Ferrenberg2003}, the free energy as a function of the reaction coordinate can be obtained. While the generator is trained using configurations from discrete windows, the network architecture and the process of training by energy allow the sampling at arbitrary bias centers (Fig.~\ref{fig:DW_data}). For this reason, an accurate free energy estimate can be obtained by increasing samples or increasing the window count even if the training data alone are not sufficient for the free energy reconstruction, as shown in Fig.~\ref{fig:DW_free_energy}. In addition, the ability to train the conditioned Boltzmann generator at different temperatures allows for the reconstruction of free energies at different values of $k_{\text{B}} T$, see Fig.~\ref{fig:DW_free_energy}. 

\subsection{Exploring Path Space}

The ability to sample independent configurations in targeted regions of phase space opens up new possibilities to investigate rare events, as generated points can serve as shooting points for trajectories. Initial tests that employ this path sampling scheme are performed on a bistable double well model. The potential energy function is constructed in a way such that two reaction channels separated by an energy barrier connect the two stable basins (Fig.~\ref{fig:model_systems}B, system parameters in SI).

With the trained network at hand, we compare three different path sampling methods: TPS using two-way shooting with randomized velocities (standard TPS)~\cite{Csajka1998}, TPS with a bias on the shooting point selection~\cite{Jung2017} and path sampling from network-generated shooting points. Shooting range TPS is included since it is the closest Markov chain-based scheme to the proposed path generation from presampled, biased shooting points. In shooting range TPS, the selection probability $p_{\text{sel}}$ of a shooting point on the previous path is biased via an arbitrary, user-defined function of a reaction coordinate. For the comparison to the network-based scheme, we use a Gaussian centered at the top of the barrier.

\begin{figure*}
    \centering
    \includegraphics[width=17cm]{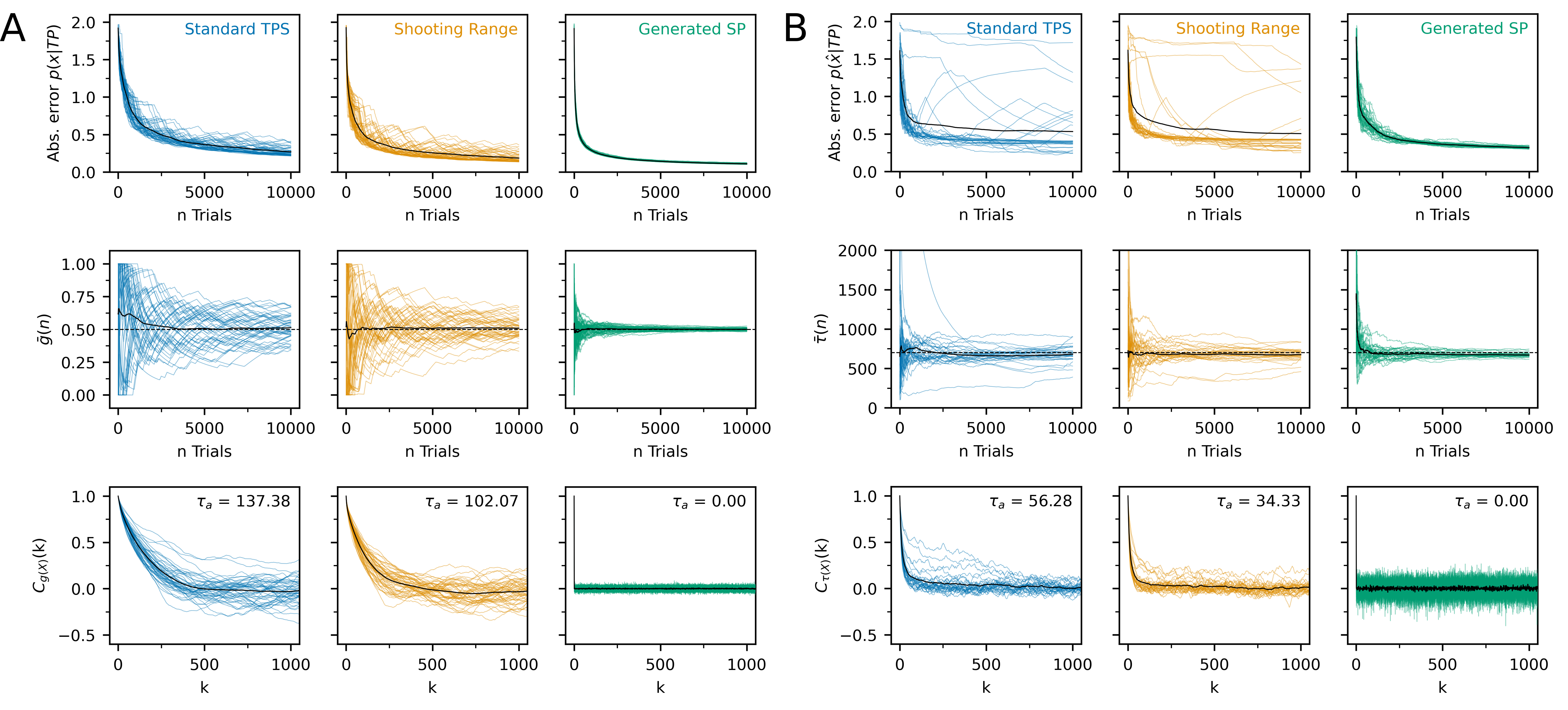}
    \caption{\textbf{Performance comparison of standard TPS (blue), shooting range TPS (orange) and sampling from generated shooting points (green) in the bistable double well model (A) and polymer model (B).} For all algorithms, sampling was performed (A) $50$ and (B) $30$ times. Each line represents a single path sampling run whereas the solid black line indicates the average over these runs. The absolute error of $p(x|\text{TP})$ as a function of the number of trials (first row) measures the deviation from a reference path ensemble at a given trial. The second row shows the running average of the reaction channel indicator function $g(X)$ in (A) and in (B) the running average of the path length $\tau$. The expected value is given by the dashed black line. As a measure of the correlation between paths, the autocorrelation function of the indicator function $g(X)$ (A) and path length (B) is shown in the third row.}
    \label{fig:TPS_performance}
\end{figure*}

The results of these initial tests show that in contrast to sampling from generated shooting points, both standard TPS and shooting range TPS struggle to estimate the ratio between paths in the upper and lower reaction channel correctly (Fig.~\ref{fig:TPS_performance}A). To obtain a quantitative measure of this ratio, we define an indicator function $g(X)$ that describes whether the path follows the upper or lower reaction channel. The function returns unity if the path follows the upper channel and vanishes otherwise. With reference to Fig.~\ref{fig:DW_data}D, the function $g(X)$ is given by
\[
g(X) = 
\begin{cases}
1 & \text{if} \ \bar{X}^y(\tau) > 0 ; \\
0 & \text{otherwise} ,
\end{cases}
\]
where $\bar{X}^y$ is the average of the $y$ component of the trajectory over the single transition path.

The expectation value of $g(X)$ over all reactive paths is $1/2$ since the dynamics and the state definitions are symmetric. For standard TPS and shooting range TPS, the average value of the indicator function oscillates around the expected value. Since correlations between sampled paths are unavoidable with shooting moves, subsequently sampled paths are likely to remain in the same reaction channel. A switch to the other channel is only observed infrequently as indicated by the integrated autocorrelation times. This leads to the oscillating behavior of the average indicator function. In comparison, the path sampling from generated shooting points produces independent paths in the upper and lower reaction channel, leading to fast convergence of the average value of the indicator function, as shown in Fig.~\ref{fig:TPS_performance}A. 

To further compare the performance of the different path sampling schemes, a reference path ensemble is sampled by means of $250{,}000$ trials using two-way shooting TPS with randomized velocities. The difference between the reference path ensemble and a path ensemble at trial $n$ can then be obtained by comparing the discretized density of configurations on transition paths $p(x|\text{TP})$ as in~\cite{Jung2017}. 
The neural-network based sampling scheme outperforms standard TPS and shooting range TPS when looking at this difference between the path ensemble at trial $n$ and the rigorously sampled reference ensemble. Moreover, due to the reweighting of generated configurations and paths, a proper distribution of paths can be obtained even if the generated shooting points are biased towards one reaction channel.

To test the scalability of the approach to higher dimensional systems, we consider a polymer model of $N=7$ beads in two dimensions, as illustrated in Fig.~\ref{fig:model_systems}C. The interaction between beads includes a non-bonded Lennard-Jones interaction, a bond stretching term and an angular term (details of the potential and simulation parameters in SI). Two stable states can be identified in this system, an extended and a compact configuration (Fig.~\ref{fig:model_systems}C). Since the states are solely identified by the radius of gyration and the Lennard-Jones interaction energy, all possible bonding permutations are included in the states. The transition between these states not only takes place infrequently but also occurs via different reaction channels making it an ideal test system for rare event sampling. 

Just like in the two-dimensional model case, we benchmark standard TPS, shooting range TPS and path sampling from network-generated shooting points on the transition from extended to circular states in the polymer model. As a coordinate for the shooting range bias and generation of initial points, we choose the radius of gyration $R_G$ of the polymer. To compare the configuration density in the path ensemble at trial $n$ with the reference ensemble for the different methods (Fig.~\ref{fig:TPS_performance}B), we discretize the configuration space of the polymer (see Materials and Methods). In contrast to sampling from generated shooting points, some standard and shooting range TPS runs show slow or non-existent convergence towards the reference path ensemble. This effect may be explained by looking at the average transition path lengths or the autocorrelation function of the path length. Path sampling runs that do not converge to the reference ensemble come with an over- or underestimated average path length. Combined with the long correlation times of the path length, it can be concluded that standard TPS and shooting range TPS are prone to get stuck in a faster or slower reaction channel compared to the typical reaction channel. 
Since sampled paths from generated shooting points are uncorrelated, all reaction channels are visited independently in proportion to their statistical weight. This leads to a consistent convergence to the reference ensemble and an accurate estimate of the average path length.

\subsection{Finding Transition States}

\begin{figure}
    \centering
    \includegraphics[width=0.7\columnwidth]{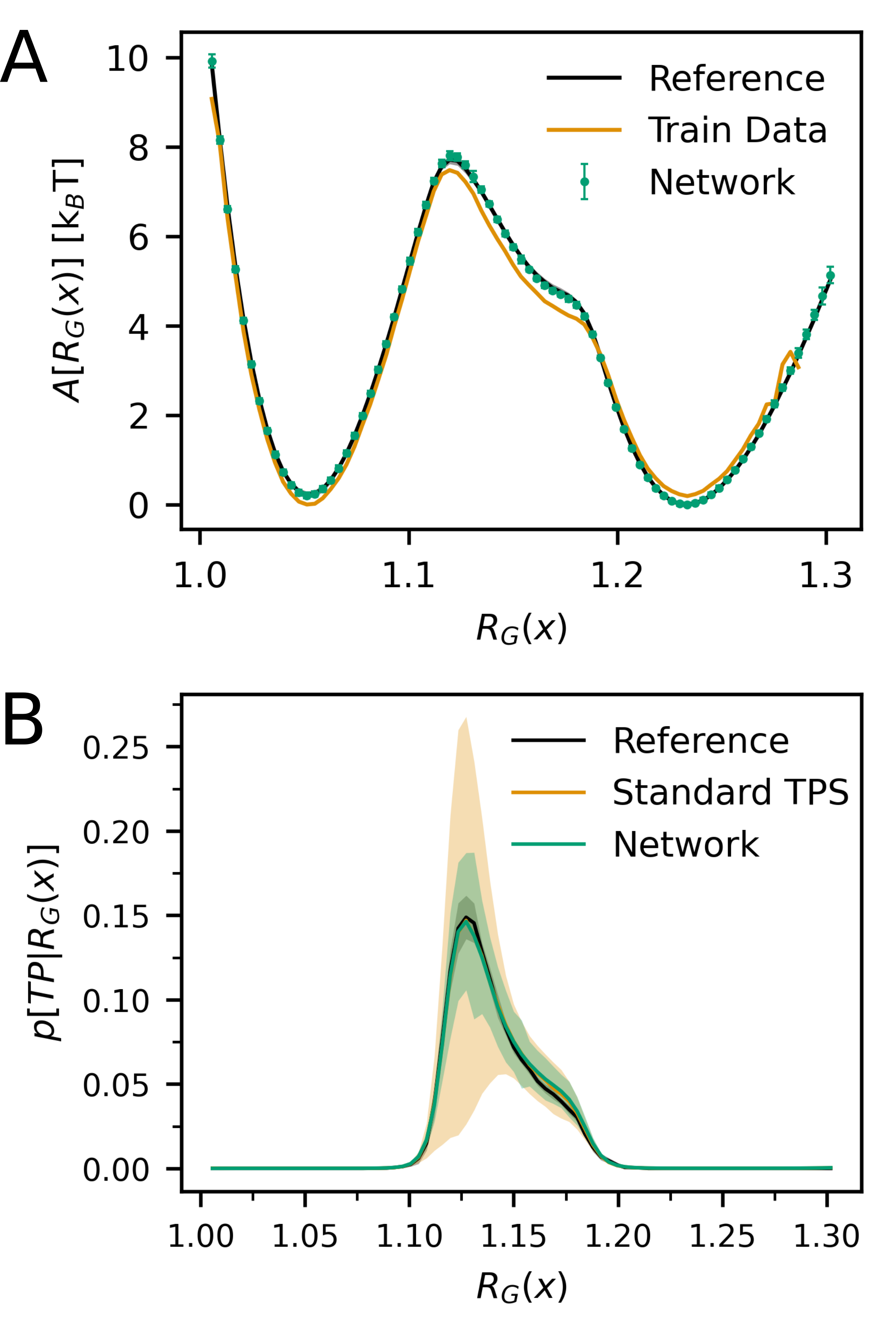}
    \caption{\textbf{Estimation of the PMF and transition path probability from network-generated configurations for the polymer model.} Comparison between the PMF reconstructed from long replica-exchange umbrella sampling runs as a reference, the training data and network-generated (upper panel). 
    The lower panel shows a comparison of the transition path probability along the radius of gyration estimated using fleeting trajectories (Reference), using standard TPS in combination with umbrella sampling (Standard TPS) and using path sampling from generated shooting points together with a network-based free energy reconstruction (Network).}
    \label{fig:polymer_PMF_pTP}
\end{figure}

The conditioning of the Boltzmann generator also allows for a closer study of the bias coordinate. A free energy profile can be reconstructed from configurations at different bias centers as already shown for the double well. This also applies to the polymer model, where the Boltzmann generator learns the correct free energy profile along the reaction coordinate even with improperly weighted training data (Fig.~\ref{fig:polymer_PMF_pTP}A).

While the position of the barrier top can give initial information on the position of possible transition states, the central quantity of interest is the transition path probability $p(\text{TP}|r(x))$, i.e. the conditional probability for a point $x$ to be on a transition path given the value of the reaction coordinate $r(x)$. Usually one obtains this measure by producing multiple fleeting trajectories from configurations with the specific reaction coordinate value~\cite{Hummer2004}. An immediate drawback of this approach is that fleeting trajectories need to be produced separately from the path sampling run, making the calculations expensive. Alternatively, one could estimate $p(\text{TP}|r(x))$ up to a proportionality constant using Bayes' theorem, as proposed by Hummer~\cite{Hummer2004}:
\begin{align}
\label{eq:pTP_Bayes}
p(\text{TP}|r(x)) \propto \frac{p(r(x)|\text{TP})}{p_{\text{eq}}(r(x))} .
\end{align}
However, since it is not trivial to extract information on the equilibrium distribution $p_{\text{eq}}(r(x))$ from the transition path ensemble, the calculation of $p(\text{TP}|r(x))$ using the Bayesian approach requires additional simulations and is therefore usually less efficient than estimation via fleeting trajectories. Moreover, both distributions $p(r(x)|\text{TP})$ and $p_{\text{eq}}(r(x))$ come with an uncertainty when estimated from simulation data and this uncertainty propagates to the estimated transition path probability.

The path sampling from network-generated shooting points proposed in this work allows for both the accurate reconstruction of the free energy and for the calculation of the path distribution along the reaction coordinate. To demonstrate this, we compare the calculation of $p(\text{TP}|R_G(x))$ for the polymer model using the Bayesian approach (Fig.~\ref{fig:polymer_PMF_pTP}B) with the results obtained using the standard approach employing fleeting trajectories. In the following comparison, we compute error estimates by performing simulations in replicas and by using Gaussian error propagation. We first estimate $p(\text{TP}|R_G(x))$ from $10$ independent runs each with $25{,}000$ fleeting trajectories as a reference. As a second baseline, we use the reference path ensemble (same as in Fig.~\ref{fig:TPS_performance}B) and the reference free energy (same as in Fig.~\ref{fig:polymer_PMF_pTP}) to estimate the transition path probability. Here the resulting error does not allow for accurate determination of transition state regions on the reaction coordinate. In comparison, using the data from the Boltzmann generator (Fig.~\ref{fig:polymer_PMF_pTP}A and the right row in Fig.~\ref{fig:TPS_performance}B) leads to a more accurate estimate of the transition path probability. The advantage of this approach is that the calculation is inexpensive since the trained network enables fast estimation of free energy profiles and efficient transition path sampling.

\section{Discussion and Conclusion}

In this work, we introduced the conditioning of Boltzmann generators for enhanced sampling of low probability regions in configuration space. The conditioned generators can be used, in the first place, to obtain more accurate free energy profiles. Secondly, we proposed a path sampling scheme based on a set of presampled, network-generated shooting points. While Boltzmann generators were a natural choice for this proof of concept, the sampling scheme including the path reweighting factor derived in Eq.~\ref{reweighting_factor_path} can be generalized to any generative machine learning scheme as long as the generated distribution of points is a good approximation to the desired shooting point distribution. Recently proposed stochastic normalizing flows~\cite{NEURIPS2020_41d80bfc}, smooth flows~\cite{NEURIPS2021_167434fa} or equivariant normalizing flows~\cite{Wirnsberger2020} can easily be adopted depending on the system to study. 

The computational cost of the potential energy evaluations for the training of the shooting point-generating network, which includes the generation of training data, is negligible compared to the cost of the trajectory propagation. This is due to the trajectory generation relying on numerous repeated, small steps each requiring a full force evaluation. As a result, there is a substantial margin for larger training sets or more expensive networks in the proposed path sampling scheme. Consequently, even though normalizing flows are known not to scale well to higher-dimensional systems (at the moment), future developments can be readily used in the proposed path sampling scheme without the need for rigorous performance benchmarks.

It is important to note that, for both use cases --- free energy reconstruction and path sampling --- we based our algorithm on a reaction coordinate $r(x)$. In the systems discussed, this coordinate is either trivial to find or could be obtained by educated guessing. Only after the training of the network and the whole path sampling process, it is possible to obtain a measure of the quality of the chosen reaction coordinate, e.g. by estimating the probability to generate a transition path. This approach is not straightaway transferable to more complex systems as reaction coordinates often turn into a less intuitive combination of order parameters~\cite{Juraszek2008,Bolhuis2000}. A direct approach to tackle this problem may be to use existing algorithms for reaction coordinate optimization such as the algorithm proposed by Peters and Trout~\cite{Peters2006} or to use a reinforcement learning scheme as proposed by Jung et al.\cite{Jung2019}. Also, even though a reaction coordinate is often challenging to find, a reasonable order parameter may sometimes be more apparent. Here the difference is that an order parameter may distinguish between different states of the system but does not necessarily have a defined, compact region linked to a high probability to generate a transition path. Therefore, as an alternative to prior reaction coordinate analysis, the functional form of the bias potential could be adapted. Instead of a harmonic bias centered on a specific region on the coordinate, one could realize the biasing via a history-dependent bias potential as employed in metadynamics~\cite{Laio2002}. With this approach, low probability states along the whole reaction coordinate could be sampled eliminating the need for centering the shooting points on a specific region.

For the future, we see the approach of using generative neural networks for rare event sampling as especially useful if prior knowledge of a reaction coordinate exists and multiple orthogonal reaction channels complicate the sampling.

\begin{acknowledgments}
We acknowledge the financial support of the Austrian Science Fund (FWF) through the SFB TACO, Grant number F 81-N. The computational results presented have been achieved in part using the Vienna Scientific Cluster (VSC).

\end{acknowledgments}

\newpage

\onecolumngrid
\normalsize
\patchcmd{\large}{15}{15}{}{}
\begin{center}
  \textbf{\LARGE Supplementary Information: Conditioning Boltzmann Generators for Rare Event Sampling}\\[.2cm]
  Sebastian Falkner,$^{1}$ Alessandro Coretti,$^{2}$ Salvatore Romano,$^{1}$ Phillip Geissler$^{3}$ and Christoph Dellago$^{2,*}$\\[.1cm]
  {\itshape ${}^1$University of Vienna, Faculty of Physics \& Vienna Doctoral School in Physics, Boltzmanngasse 5, A-1090 Vienna, Austria.\\
  ${}^2$University of Vienna, Faculty of Physics, 1090, Vienna, Austria.\\
  ${}^3$Department of Chemistry, University of California, Berkeley, California 94720, USA\\
  }
  ${}^*$Electronic address: christoph.dellago@univie.ac.at\\
(Dated: \today)\\[2cm]
\end{center}

\setcounter{equation}{0}
\setcounter{figure}{0}
\setcounter{table}{0}
\setcounter{page}{1}
\setcounter{section}{0}
\renewcommand{\theequation}{S\arabic{equation}}
\renewcommand{\thefigure}{S\arabic{figure}}
\renewcommand{\thetable}{S\arabic{table}}
\renewcommand{\bibnumfmt}[1]{[S#1]}
\renewcommand{\citenumfont}[1]{S#1}
\renewcommand{\thesection}{S\Roman{section}}
\renewcommand{\thepage}{S\arabic{page}}

\titleformat*{\section}{\Large\bfseries}
\section{Derivation of the Path Reweighting Factor}
A given trajectory that was generated from a previously sampled shooting point can theoretically be generated from any point on the trajectory by means of a shooting move. Therefore, the total generation probability is the sum of the independent generation probabilities from each point on the trajectory:
\begin{align}
   \label{dynamical_path_probability_from_sampled_SP}
   P_{\text{gen}}\left[ X(\tau) \right]\, \text{d}X^{\tau} = & \text{d}X^{\tau}\,
   \sum_{k=0}^{\tau/\Delta t} \biggl[  p_{\text{SP}}(x_{k\Delta t})
   \bigl[h_{\text{A}}(x_0) + h_{\text{B}}(x_0)\bigr] \bigl[h_{\text{A}}(x_\tau) + h_{\text{B}}(x_\tau)\bigr]
   \prod_{j=1}^{\tau/\Delta t-1} \widetilde{h}(x_{j \Delta t}) \notag \\
   &\times\prod_{i=k}^{\tau/\Delta t-1} p(x_{i\Delta t} \to x_{(i+1)\Delta t}) 
   \prod_{i=1}^{k} \bar{p}(x_{i\Delta t} \to x_{(i-1)\Delta t}) \biggr] .
\end{align} 
The short-time transition probabilities obey detailed balance:
\begin{align}
    \label{transition_detailed_balance}
    \bar{p}(x_{i\Delta t} \to x_{(i-1)\Delta t}) = 
        p( x_{(i-1)\Delta t} \to x_{i\Delta t}) 
        \frac{p_{\text{eq}}(x_{(i-1)\Delta t})}{p_{\text{eq}}(x_{i\Delta t})} .
\end{align} 
By applying Eq.~\ref{transition_detailed_balance} repeatedly, we can rewrite Eq.~\ref{dynamical_path_probability_from_sampled_SP} as:
\begin{align}
\label{S_generation_probability_from_psp_full}
P_{\text{gen}}\left[ X(\tau) \right]\, \text{d}X^{\tau} = &\text{d}X^{\tau}\,
    \sum_{k=0}^{\tau/\Delta t} \Bigg[ p_{\text{SP}}(x_{k\Delta t})
    \frac{p_{\text{eq}}(x_{0})}{p_{\text{eq}}(x_{k\Delta t})}
    \times \bigl[h_{\text{A}}(x_0) + h_{\text{B}}(x_0)\bigr] \bigl[h_{\text{A}}(x_\tau) + h_{\text{B}}(x_\tau)\bigr]\notag\\
    &\times \prod_{j=1}^{\tau/\Delta t-1} \widetilde{h}(x_{j \Delta t})
    \prod_{i=0}^{\tau/\Delta t-1} p(x_{i\Delta t} \to x_{(i+1)\Delta t})  \Bigg] .
\end{align} 
Using the reactive path probability $P_{\text{AB}} \left[ X(\tau) \right]$ (Eq.~1 in main text), we can simplify this expression to:
\begin{align}
\label{S_generation_probability_from_psp}
P_{\text{gen}}\left[ X(\tau) \right] \, \text{d}X^{\tau} =& \, \text{d}X^{\tau} \, 
    P_{\text{AB}}\left[ X(\tau) \right]\,Z_{\text{AB}} 
    \sum_{k=0}^{\tau/\Delta t}
        \frac{p_{\text{SP}}(x_{k\Delta t}) }{p_{\text{eq}}(x_{k\Delta t})} + \parbox{2.9cm}{contributions from unreactive paths} .
\end{align} 
The contributions from the unreactive paths are not considered in the final generation probability as trajectories that start and end in the same state are anyway discarded. The weight of reactive paths in Eq.~\ref{S_generation_probability_from_psp} differs from the reactive path probability by the factor:
\begin{align}
\label{S_reweighting_factor_path}
w \left[ X(\tau) \right] = 
    \frac{1}
    {Z_{\text{AB}}} 
    \left[\sum_{k=0}^{\tau/\Delta t} \frac{p_{\text{SP}}(x_{k\Delta t}) }{p_{\text{eq}}(x_{k\Delta t})} \right]^{-1} .
\end{align}

\section{Derivation of the Training Loss Functions}

The starting point for the derivation of a loss function for the Boltzmann generators in the conditioned case is the conditional KL divergence defined, in general, as:
\begin{align}
\label{eq:general_KL_div}
	&\text{KL}\left[p(x | c ) || q(x | c)\right] 
	= \int \dd c \ p(c)  \int \dd x \ p(x | c ) \ln \frac{p(x | c )}{q(x | c )}\notag\\
	&= \int \dd c \ p(c) \left[ - H_p(c) - \int \dd x \ p(x | c ) \ln q(x | c ) \right]\notag\\
	&= \mathbf{E}_{c\sim p(c)} \left[ - H_p(c) - \int \dd x \ p(x | c ) \ln q(x | c ) \right] ,
\end{align}
where $H_p(c)$ is the entropy of the distribution $p$ given a realization of the condition $c$. In our case, we set the condition vector $c$ to be the bias center $\bar{r}$ of a harmonic potential.

From there on, the training loss for training by example can be derived based on the conditional KL-divergence between the reference and generated data distribution $\text{KL}\left[p_\text{x}(x | r ) || q_\text{x}(x | r; \theta)\right]$:
\begin{align}
\text{KL}&[p_\text{x}(x | r ) || q_\text{x}(x | r; \theta)] = \mathbf{E}_{\bar{r} \sim p(r)} \biggl[ - H_\text{x}(\bar{r}) - \int \dd x \  p_\text{x}(x | \bar{r} ) \ln q_\text{x}(x | \bar{r}; \theta) \biggr]\notag\\
&= \mathbf{E}_{\bar{r} \sim p(r)} \biggl\{ - H_\text{x}(\bar{r}) - \int \dd x \ p_\text{x}(x | \bar{r}) \Bigl[ \ln p_\text{z}( f\inv(x | \bar{r}; \theta) | \bar{r} ) + \ln \lvert \det J\inv(x | \bar{r}; \theta) \rvert \Bigr] \biggr\}\notag\\
&=  \mathbf{E}_{\bar{r} \sim p(r)} \biggl\{ - H_\text{x}(\bar{r}) - \mathbf{E}_{x \sim  p_\text{x}(x | \bar{r})} \Bigl[ \ln p_\text{z}( f\inv(x | \bar{r}; \theta) | \bar r ) + \ln \lvert \det J\inv(x | \bar{r}; \theta) \rvert \Bigl]\biggr\} \notag\\
&=  \mathbf{E}_{\bar{r} \sim p(r)} \biggl\{ - H_\text{x}(\bar{r}) + \ln Z_\text{z} + \mathbf{E}_{x \sim  p_\text{x}(x | \bar{r})} \bigg[\frac{1}{\sigma^2} || f\inv(x | \bar{r}; \theta) || ^2 - \lvert \det J\inv(x | \bar{r}; \theta) \rvert \biggr] \biggr\}\notag\\
&=  \mathbf{E}_{\bar{r} \sim p(r)}  \bigl\{ - H_\text{x}(\bar{r}) + \ln Z_\text{z} \bigr\}
+ \mathbf{E}_{\bar{r} \sim p(r)} \biggl\{ \mathbf{E}_{x \sim  p_\text{x}(x | \bar{r})} 
\biggl[\frac{1}{\sigma^2} || f\inv(x | \bar{r}; \theta) || ^2 - \lvert \det J\inv(x | \bar{r}; \theta) \rvert \biggr] \biggr\} .
\end{align}

In the reverse direction, the conditional KL-divergence between the reference and generated latent distribution $\text{KL}\left[p_\text{z}(z | r ) || q_\text{z}(z | r; \theta)\right] $ corresponds to the training loss:
\begin{align}
\text{KL}&[p_\text{z}(z | r ) || q_\text{z}(z | r; \theta)] = \mathbf{E}_{\bar{r} \sim p(r)} \biggl[ - H_\text{z}(\bar{r}) - \int  \dd z \ p_\text{z}(z | \bar{r} ) \ln q_\text{z}(z | \bar{r}; \theta) \biggr]\notag\\
&=  \mathbf{E}_{\bar{r} \sim p(r)} \biggl\{ - H_\text{z}(\bar{r}) - \int \dd z \ p_\text{z}(z | \bar{r}) \biggl[ \ln p_\text{x}( f(z | \bar{r}; \theta) | \bar{r} ) + \ln \lvert \det J(z | \bar{r}; \theta) \rvert \biggr]  \biggr\} \notag\\
&= \mathbf{E}_{\bar{r} \sim p(r)} \biggl\{ - H_\text{z}(\bar{r}) - \mathbf{E}_{z \sim  p_\text{z}(z | \bar{r})} \biggl[ \ln p_\text{x}( f(z | \bar{r}; \theta) | \bar r ) + \ln \lvert \det J(z | \bar{r}; \theta) \rvert \biggr]\biggr\} \notag\\
&= \mathbf{E}_{\bar{r} \sim p(r)} \biggl\{ - H_\text{z}(\bar{r}) + \mathbf{E}_{z \sim  p_\text{z}(z | \bar{r})} \biggl[\beta U(f(z | \bar{r}; \theta)) + \beta \frac{k}{2} [r(f(z | \bar{r}; \theta)) - \bar{r}]^2+ \ln Z_{\text{x}} - \lvert \det J(z | \bar{r}; \theta) \rvert \biggr] \biggr\} \notag\\
&=\mathbf{E}_{\bar{r} \sim p(r)} \biggl\{ - H_\text{z}(\bar{r}) + \ln Z_{\text{x}} \biggr\} + \mathbf{E}_{\bar{r} \sim p(r)}  \biggl\{ \mathbf{E}_{z \sim  p_\text{z}(z | \bar{r})} \biggl[\beta U(f(z | \bar{r}; \theta))+ \beta \frac{k}{2} [r(f(z | \bar{r}; \theta)) - \bar{r}]^2 - \lvert \det J(z | \bar{r}; \theta) \rvert \biggr] \biggr\} .
\end{align}

The expectancy values of  $- H_\text{x}(\bar{r}) + \ln Z_\text{z}$ and $- H_\text{z}(\bar{r}) + \ln Z_{\text{x}}$ are independent of the network parameters and can therefore be omitted during training.

\section{Network Parameters and Training}

The network architecture used in this work is based on RealNVP networks~\cite{SDinh2017} as also used in the unconditioned Boltzmann generator~\cite{SNoe2019}. We use a single feed-forward network with $\tanh$-activation functions to obtain $S$ and $T$ parameters given an input $x_{1/2}$ and a condition $c$. The network parameters used for the double well model, bistable double well model, and the polymer model are given in Tab.~\ref{tbl:Network_Parameters}. Here, $N_{\text{blocks}}$ denotes the number of RealNVP blocks. Each block has a parameter network consisting of a number of hidden layers $N_{\text{layers}}$ with a certain number of nodes $N_{\text{hidden}}$. The sum of all trainable parameters is given by $N_{\text{param}}$. 

\begin{figure}
    \centering
    \includegraphics[width=13cm]{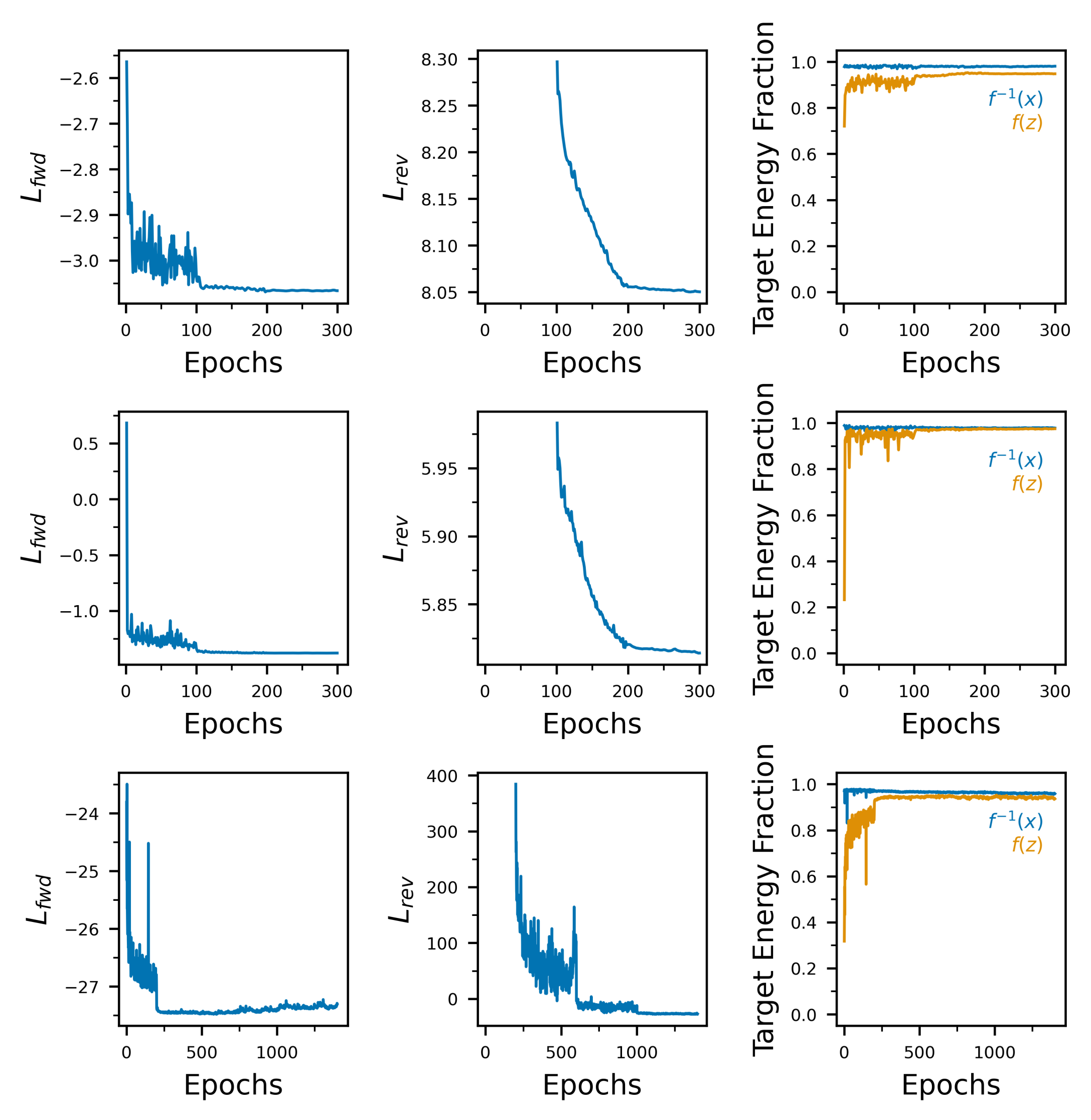}
    \caption{Training metrics for the double well model (top row), bistable double well model (middle row), and the polymer model (bottom row). The left column shows the training by example loss, the middle column the training by energy loss and the last column the fraction of generated data within the target energy range. A point is assumed to be within the target energy range if it is within the 1st and 99th energy percentile of separately sampled reference data.}
    \label{fig:train_curves}
\end{figure}

The corresponding training protocols are specified in tables~\ref{tbl:DW_train_protocol},~\ref{tbl:BSDW_train_protocol} and~\ref{tbl:Polymer_train_protocol}. The batch size, learning rate (LR), training by energy weight ($\lambda_{\text{rev}}$) and clamping start of the potential energy ($U_{\text{clamp}}$) were adapted in each stage of the training. In contrast, the parameters for the estimation of the training by example loss remain constant. In particular, the weight for the example loss is $\lambda_{\text{fwd}} = 1$ for the whole training. The number of conditions sampled in each batch and their range  ($N_{\text{cond}}$, RC Range) and the training temperature range are likewise constant. The training metrics are summarized in Fig.~\ref{fig:train_curves}.

\section{2D System Definitions}
\subsection{Double Well Model}
The two-dimensional model system (Figure~3A) is defined using the potential energy function~\cite{SJung2017}:
\begin{align}
U(x) = 10\bigl[\bigl((x^{(0)})^2 - 1\bigr)^2 + \bigl(x^{(0)}-x^{(1)}\bigr)^2\bigr] ,
\end{align} 
where $x = (x^{(0)}, x^{(1)})$. The reaction coordinate is given by $r(x) = x^{(0)} + x^{(1)}$. To train the network, $1,500$ configurations in eight linearly spaced windows between $r(x) = -3$ and $r(x) = 3$ are sampled using Monte Carlo sampling with a force constant of $k=25$ and $k_{\text{B}}T = 1$. As a reference for the free energy profile along $r(x)$, $10,000$ configurations in $30$ linearly spaced umbrella windows between $r(x) = -3$ and $r(x) = 3$ are included leading to a total of $3\times10^5$ configurations.

\subsection{Bistable Double Well Model} 
The potential energy of the bistable double well model (Figure~3B) used for initial transition path sampling tests is defined as:
\begin{align}
U(x) = \frac{15}{8}\biggl[\frac{\bigl((x^{(0)})^2  + (x^{(1)})^2 - 4\bigr)^2}{4} + (x^{(1)})^2\biggr] .
\end{align} 
We use $r(x) = x^{(0)}$ as the bias coordinate for the training of the Boltzmann Generator. For the generation of the training data, we use a force constant of $k=8$ and sample $1,500$ configurations in six uniformly spaced windows between $r(x) = -3$ and $r(x) = 3$ at $k_{\text{B}}T = 1$. Transition path sampling is performed using an underdamped Langevin integrator~\cite{SGoga2012} with $\gamma=20$, $\Delta t=10^{-2}$ and $k_{\text{B}}T = 1$. The system is assumed to be in state A when $(x^{(0)}-2.2)^2 + (x^1)^2 < 0.1$ and in state B when $(x^{(0)}+2.2)^2 + (x^1)^2 < 0.1$. To compare the performance of different path sampling schemes, a reference path ensemble is sampled by means of $250,000$ trials using two-way shooting TPS with randomized velocities. For shooting range TPS, we use a Gaussian bias of the form:
\begin{align}
p_{\text{sel}}[ r(x) ] \propto \exp\bigl\{ - \zeta [ r(x) - \mu ]^2 \bigr\} ,
\end{align}
where $\zeta =12.5$ and $\mu = 0$ for the bistable double well model. 

The sampling from generated shooting points is initiated by generating points and resampling them with a weight according to~\cite{SNoe2019}:
\begin{align}
\label{BG_reweighting_factor}
\omega(x) \propto \exp \Bigl\{ -U(f(z | \bar{r}; \theta))
+ \frac{1}{\sigma^2} || z || ^2 +  \ln \lvert \det J(z | \bar{r}; \theta) \rvert \Bigr\} .
\end{align}
The path reweighting factor (equation~\ref{S_reweighting_factor_path}) for a harmonic bias on the shooting points reduces to:
\begin{align}
\label{reweighting_factor_reduces}
\omega\left[ X(\tau) \right] \propto
\left[\sum_{k=0}^{\tau/\Delta t} \exp\biggl[-\beta \frac{k}{2} \bigl(r(x_{k\Delta t}) - \bar{r}\bigr)^2\biggr] \right]^{-1} .
\end{align} 
The difference between the reference path ensemble and the path ensemble at trial $n$ is obtained by comparing the density of configurations on transition paths $p(x|\text{TP})$. We compute the histogram of the configurations obtained using each sampling scheme and sum the absolute probability density difference for each bin referred to by the indices $i$ and $j$:
\begin{align}
\text{Abs. Error }p(x|\text{TP}) = \sum_{i, j} | p(x_{ij}|\text{TP}) - p_{\text{ref}}(x_{ij}|\text{TP})| .
\end{align}

\section{Polymer Model}

The polymer model of $N=7$ beads in two dimensions is defined by pairwise interaction between beads including a nonbonded Lennard-Jones interaction, a bond stretching term and an angular term:
\begin{align}
U_{\text{LJ}} &= \sum_{i=1}^N \sum_{j>i}^{N} 4\epsilon \left[ \left(\frac{\sigma}{d_{ij}}\right)^{12} - \left(\frac{\sigma}{d_{ij}}\right)^{6}\right] , \\
U_{\text{bond}} &= \sum_{i=1}^{N-1} \frac{k_{\text{bond}}}{2} (d_{i,i+1} - d_{\text{ref}})^2 , \\
U_{\text{angle}} &= \sum_{i=1}^{N-2} \frac{k_{\text{angle}}}{2} [1 - \cos(\varphi_{i,i+1,i+2} - \varphi_{\text{ref}})] ,
\end{align}
where $d_{ij}$ is the distance between atoms $i$ and $j$, $\sigma = 1$, $\epsilon = 1$,  $k_{\text{bond}} = 5\,\epsilon \sigma^{-1}$,  $d_{\text{ref}} = \sqrt[6]{2}\,\sigma$, $k_{\text{angle}} = 1.4\,\epsilon\, \text{rad}^{-1}$ and $\varphi_{\text{ref}} = \pi \, \text{rad}$. The angle $\varphi$ is defined between $0$ and $2 \pi$.

The extended and circular states of the model are given by $U_{\text{LJ}} < -11.2\,\epsilon$ and $R_G(x) > 1.2\,\sigma$ and $U_{\text{LJ}} < -12.3\,\epsilon$ and $R_G(x) < 1.05\,\sigma$ respectively, where $R_G(x)$ is the radius of gyration of the polymer chain.

The training data are obtained using replica exchange umbrella sampling. As a bias coordinate in this system, we chose the radius of gyration:
\begin{align}
R_G(x) = \sqrt{ \frac{1}{N} \sum_{i=1}^{N} (d^{\text{com}}_i)^2} ,
\end{align}
where $d^{\text{com}}_i$ is the distance to the center of mass of the $i$-th bead of the polymer. Configurations are sampled in $10$ uniformly spaced windows between $R_G(x)=1.013\,\sigma$ and $R_G(x)=1.25\,\sigma$ with a force constant of $1000\, \epsilon\sigma^{-1}$ at $k_{\text{B}} T=0.1\,\epsilon$. In each window, $5\times10^6$ MC steps are performed while attempting an exchange between neighboring windows every $500$ steps. Every $250$ steps a configuration is saved for training leading to $20,000$ configurations in each window. Reference data for the free energy calculations are obtained by $1\times10^{7}$ MC steps in $24$ uniformly spaced windows between $R_G(x)=1\,\sigma$ and $R_G(x)=1.3\,\sigma$ attempting an exchange every $1000$ steps.

Transition path sampling is performed as in the double well model using an underdamped Langevin integrator with $\gamma=4\,(m \sigma^2 / \epsilon)^{-\frac{1}{2}}$, $\Delta t=5\times10^{-3}\,\sqrt{m \sigma^2 / \epsilon}$ and $k_{\text{B}}T = 0.1 \epsilon$. The extended and compact state definitions of the model can be found in the supplementary information. A reference ensemble is produced by means of $15,000$ trials with $20$ independent walkers summing up to $3\times10^{5}$ trials. For shooting range TPS, we use the bias parameters $\sigma = \frac{1000}{2}\, \epsilon\sigma^{-1}$ and $\mu = 1.125\, \sigma$. Path sampling from generated shooting points is initiated from a pool of $10^{6}$ points which are resampled according to the weight described in Eq.~\ref{BG_reweighting_factor}. In contrast to the two-dimensional model, the density of configurations in the polymer transition path ensemble is not obtainable via binning due to the higher dimensionality. We solve this by discretizing configuration space described more detailed in the next section.

For the training and generation process, the system is represented in internal coordinates defined via the distances and angles between the beads so that the network does not have to learn the translational and rotational symmetries of the system. The determinant of the transformations Jacobian required for the evaluation of the loss function is given by:
\begin{align}\label{logdet}
    \log \det J_{\text{int}}(a) = \sum_{i=2}^{N-1} \log d_{i-1,i} ,
\end{align}
where $a$ describes a set of internal coordinates and $d_{ij}$ is the distance between bead $i$ and $j$. A full derivation is provided in the supplemental information. We removed mirror images in the training set by constraining the first angle to $[0, \pi)$ and flipping every molecule with a first angle outside this range. The resulting internal coordinates were normalized before entering the network by subtracting the training set's mean and dividing by its standard deviation.

\section{Discretization of the Polymer Model Configuration Space}

\begin{figure}
    \centering
    \includegraphics[width=13cm]{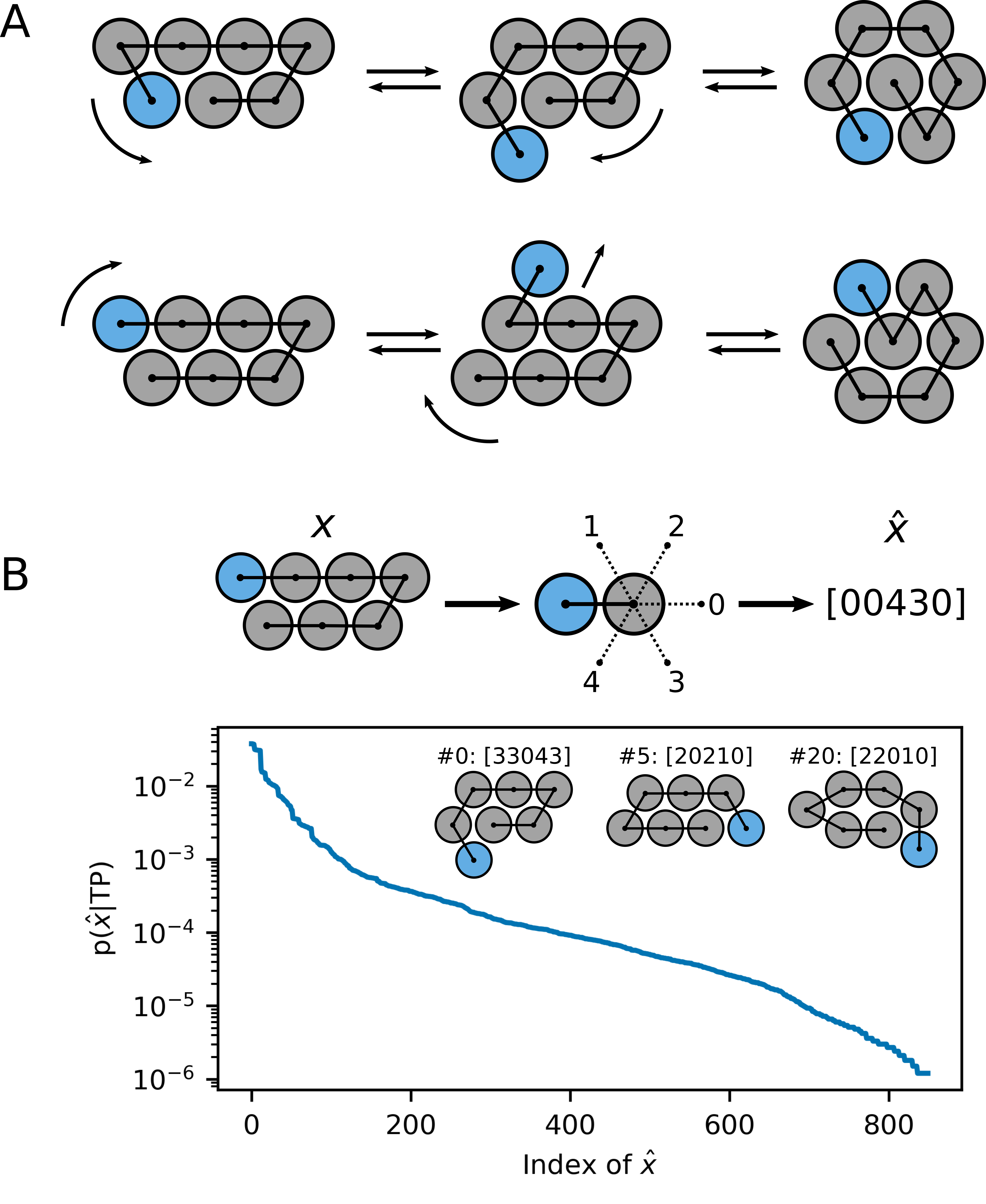}
    \caption{\textbf{Rare events in the polymer model and discretization of the configuration space.} (A) The polymer model consisting of seven beads can undergo a transition from an extended to a circular conformation. While the upper path sketches the most common mechanism observed in the transition path ensemble, alternative pathways such as the lower one occur occasionally. (B) The configuration space is discretized by assigning each angle of the polymer molecule a number between zero and four. This allows to model the probability of a configuration on a transition path using a discrete probability density function $p(\hat{x}|\text{TP})$ (lower panel).}
    \label{fig:polymer_description}
\end{figure}

 We estimate the densities of configurations on transition paths for the polymer system by discretizing the configuration space (Fig.~\ref{fig:polymer_description}B). The five angles of the polymer model were each assigned a class based on the nearest multiple of $60^\circ$. Bond distances were neglected in the discretized model as their small fluctuations are irrelevant for the definition of the state. Applying this procedure, one ends up with a five-digit number describing the turns of the polymer chain. A total of $5^5 = 3125$ discretized configurations are theoretically possible. However, many of them occur with vanishing probability due to particle overlap. This effect is observable in the reference transition path ensemble that is produced by means of $15{,}000$ standard TPS trials each in $20$ independent walkers summing up to $3\times10^{5}$ trials. In this reference encompassing $209$ million configurations, we observe approximately $800$ unique discrete configurations (Fig.~\ref{fig:polymer_description}B). Only approximately $100$ of them occur more frequently than 1 in 1000.
 
\section{Internal Coordinate Transformation for the Polymer Model}

Training the Boltzmann generator on an internal coordinate representation requires the calculation of the Jacobian of the transformation in order to evaluate the loss function.

We begin by representing the polymer as a chain of $N$ atoms in two-dimensional space. The $i$-th atom position ($i=0, \ 1, \dots, N-1$) in Cartesian coordinates is denoted by $x_i=(x^{(0)}_i,x^{(1)}_i)$. The reference frame is chosen so that the first atom coincides with the origin and the second atom lies along the $x^{(0)}$-axis, i.e. $x^{(0)}_0=x^{(1)}_0=0$ and $x^{(1)}_1=0$. In this way, the rotation and translation of the molecule is fixed. Thus, we are left with $2N-3$ degrees of freedom that are represented as a vector of Cartesian coordinates $x=(x^{(0)}_1;x^{(0)}_2,x^{(1)}_2;\dots; x^{(0)}_i,x^{(1)}_i;\dots; x^{(0)}_{N-1},x^{(1)}_{N-1})$.

\begin{figure}
    \centering\includegraphics[width=10cm]{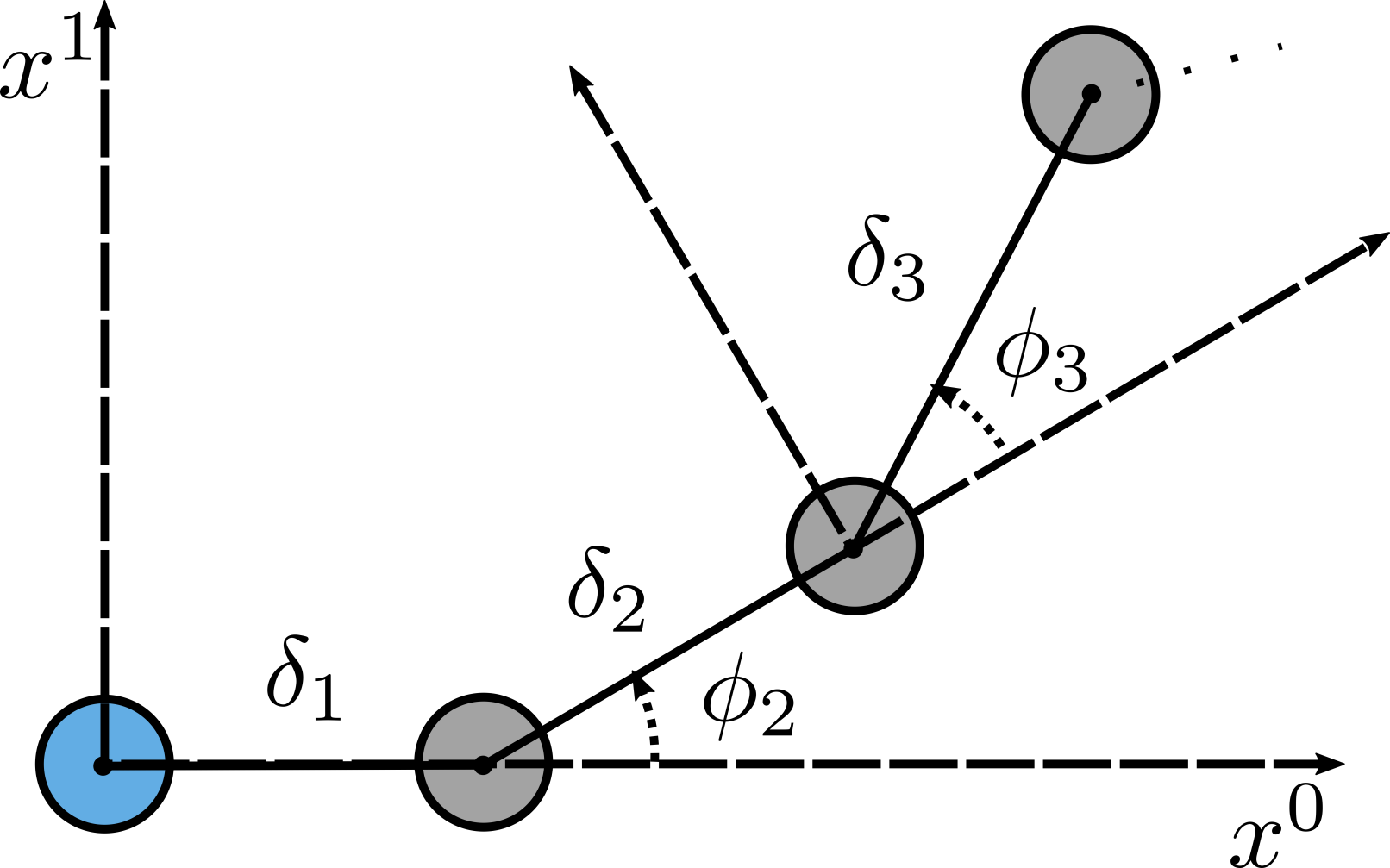}
    \caption{Representation of a generic polymer in two dimensions.}
    \label{fa}
\end{figure}

The polar coordinates (Fig.~\ref{fa}) are defined by:
\begin{itemize}
    \item the bond length $\delta_i$ is the distance between atom $i-1$ and atom $i$.
    \item the angle $\phi_i$ is the angle between atoms $i-2$, $i-1$ and $i$
\end{itemize}
The choice of the reference frame in Cartesian coordinates leads to $2N-3$ polar degrees of freedom: $a=(\delta_1;\delta_2,\phi_2;\dots; \delta_i,\phi_i;\dots; \delta_{N-1},\phi_{N-1})$.
The derivation of an exact expression of the determinant of the Jacobian makes use of the inverse transformation $x=x(a)$, which can be written as
\begin{equation}
\label{inverse_internal_transformation}
x_i=x_{i-1} + \delta_i R(\phi_i)R(\phi_{i-1})\dots R(\phi_2)\begin{bmatrix} 1\\  0\\  \end{bmatrix} ,
\end{equation}
where $R(\phi)$ is the matrix representing the rotation by an angle $\phi$.
Using the property of two-dimensional rotations $\prod_k R(\phi_k)=R(\sum_k \phi_k)$, \eqref{inverse_internal_transformation} can be written as:
\begin{equation}
 \begin{cases}
x^{(0)}_i=x^{(0)}_{i-1}+\delta_i\cos(\phi_{i}+\phi_{i-1}+\dots+ \phi_2) , \\
x^{(1)}_i=x^{(1)}_{i-1}+\delta_i\sin(\phi_{i}+\phi_{i-1}+\dots+ \phi_2) . \\
\end{cases}   
\end{equation}
The position of an atom in the polymer chain depends only on the coordinates of the previous particles. Hence, the derivative of the Cartesian coordinates with respect to all the following polar coordinates is trivially zero. This means that the Jacobian of the transformation is a lower-triangular block matrix. As a result, the determinant can be computed as the product of the determinants of the blocks on the diagonal $J_i$, i.e. 
\begin{equation}\label{b}
    \det J=\prod_{i=2}^{N-1} \det J_i , 
\end{equation}
where
\begin{equation}
   J_i \equiv \frac{\partial (x^{(0)}_i,x^{(1)}_i)}{\partial (\delta_i,\phi_i)}=\begin{bmatrix}
   \cos(\phi_{i}+\cdots \phi_2) & -\delta_i\sin(\phi_{i}+\cdots \phi_2)\\
   \sin(\phi_{i}+\cdots \phi_2) & \delta_i \cos(\phi_{i}+\cdots \phi_2)\\
   \end{bmatrix} ,
\end{equation}
and whose determinant is given by $\det J_i = \delta_i$. Therefore, the determinant of the Jacobian of the whole transformation is given by:
\begin{equation}
    \det J =\prod_{i=2}^{N-1} \delta_i= \delta_2 \delta_3\dots \delta_{N-1} .
\end{equation}

\begin{table}
   \centering
  \caption{Network Parameters for the double well model (DW), bistable double well model (BSDW), and the polymer model. }
  \label{tbl:Network_Parameters}
  \begin{tabular}{lrrrr}
    \hline
    System & $N_{\text{blocks}}$  & $N_{\text{layers}}$ & $N_{\text{hidden}}$ & $N_{\text{param}}$\\
    \hline
    DW & 3 & 3 & 64 & $26,892$\\
    BSDW & 4 & 3 & 100 & $84,816$\\
    Polymer & 8 & 3 & 200 & $302,256$\\
    \hline
  \end{tabular}
\end{table}

\begin{table}
\centering
  \caption{Training protocol for the double well model.}
  \label{tbl:DW_train_protocol}
  \begin{tabular}{rrrrrrrr}
    \hline
    Epochs & LR & Batch Size & $\lambda_{\text{rev}}$ & $N_{\text{cond}}$  & T & RC Range & $U_{\text{clamp}}$ \\
    \hline
    100 & 0.01 & 128 & 0 & - & 0.5-5 & - & -\\
    100 & 0.001 & 2500 & 1 & 50 & 0.5-5 & $-3$ - $3$ &  $10^6$\\
    100 & 0.0001 & 2500 & 1 & 50 & 0.5-5 & $-3$ - $3$ & $10^4$\\
    \hline
  \end{tabular}
\end{table}

\begin{table}
\centering
  \caption{Training protocol for the bistable double well model.}
  \label{tbl:BSDW_train_protocol}
  \begin{tabular}{rrrrrrrr}
    \hline
    Epochs & LR & Batch Size & $\lambda_{\text{rev}}$ & $N_{\text{cond}}$  & T & RC Range & $U_{\text{clamp}}$ \\
    \hline
    100 & 0.01 & 128 & 0 & - & 1 &  - & -\\
    100 & 0.001 & 2500 & 1 & 50 & 1 &  $-3$ - $3$ & $10^6$\\
    100 & 0.0001 & 2500 & 1 & 50 & 1 &  $-3$ - $3$ & $10^4$\\
    \hline
  \end{tabular}
\end{table}

\begin{table}
    \centering
  \caption{Training protocol for the polymer model.}
  \label{tbl:Polymer_train_protocol}
  \begin{tabular}{rrrrrrrr}
    \hline
    Epochs & LR & Batch Size & $\lambda_{\text{rev}}$ & $N_{\text{cond}}$  & T & RC Range & $U_{\text{clamp}}$ \\
    \hline
    200 & 0.001 & 128 & 0 & - & 0.05-0.3& - & -\\
    400 & 0.0001 & 2500 & 0.0001 & 50 & 0.05-0.3 & $1\, \sigma$ - $1.25\, \sigma$ & $5\times10^4$\\
    400 & 0.0001 & 2500 & 0.001 & 50 & 0.05-0.3 & $1\, \sigma$ - $1.25\, \sigma$ & $10^4$\\
    400 & 0.0001 & 2500 & 0.01 & 50 & 0.05-0.3 & $1\, \sigma$ - $1.25\, \sigma$ & $10^4$\\
    \hline
  \end{tabular}
\end{table}

\end{document}